\documentclass[aps,prl,twocolumn,showpacs]{revtex4}
\usepackage{graphicx,bm}
\begin{document}
\newcommand{\half}{\frac{1}{2}}
\newcommand{\ith}{^{(i)}}
\newcommand{\im}{^{(i-1)}}
\newcommand{\gae}
{\,\hbox{\lower0.5ex\hbox{$\sim$}\llap{\raise0.5ex\hbox{$>$}}}\,}
\newcommand{\lae}
{\,\hbox{\lower0.5ex\hbox{$\sim$}\llap{\raise0.5ex\hbox{$<$}}}\,}

\title{Exact characterization of O($n$) tricriticality in two dimensions}
\author{Wenan Guo~$^{1}$, Bernard Nienhuis$^{2}$ and
Henk W.J. Bl\"ote~$^{3,4}$}
\affiliation{$^{1}$Physics Department, Beijing Normal University,
Beijing 100875, P. R. China }
\affiliation{$^{2}$Instituut voor Theoretische Fysica,
Universiteit van Amsterdam, Valckenierstraat 65, The Netherlands}
\affiliation{$^{3}$Faculty of Applied Sciences, Delft University of
Technology, P. O. Box 5046, 2600 GA Delft, The Netherlands}
\affiliation{$^{4}$ Instituut Lorentz, Leiden University,
  P.O. Box 9506, 2300 RA Leiden, The Netherlands}

\date{\today} 
\begin{abstract}
We propose exact expressions for the conformal anomaly and for three  
critical exponents of the tricritical O($n$) loop model as a function
of $n$ in the range $-2 \leq n \leq 3/2$. These findings are based on
an analogy with known relations
between Potts and O($n$) models, and on an exact solution of a
'tri-tricritical' Potts model described in the literature. We verify
the exact expressions for the tricritical O($n$) model by means of a 
finite-size scaling analysis based on numerical transfer-matrix
calculations.
\end{abstract}
\maketitle 

While exact results exist for a rich collection of universality classes 
of two-dimensional phase transitions, including $q$-state Potts
criticality and tricriticality, and O($n$) criticality,
such results are still absent for the tricritical O($n$) model,
except for isolated points at $n=0$ and 1. The purpose of the present
work is to fill in this gap and to provide exact formulas for the
conformal anomaly and the main scaling dimensions of the tricritical
O($n$) model as a function of $n$. These results are not rigorous in
the mathematical sense but they may still be assumed to be exactly
true, as we shall argue below. 

The O($n$) model is defined in terms of $n$-component spins on a 
lattice, with an isotropic pair coupling of the form
$E_{ij}= \epsilon({\vec{S}}_i \cdot {\vec{S}}_j)$ where $i$ and $j$
denote two neighboring lattice sites and $\epsilon$ is a function. 
This model can be represented by a graph expansion \cite{Stanley},
in which $n$ assumes the role of a continuous parameter.
For this purpose, it is especially useful to choose the model on
the honeycomb lattice, and the function $\epsilon$ as
$\epsilon(p)\equiv - \log(1+xp)$, where $x$ is a measure of the
inverse temperature. Then the graph expansion reduces to a gas
of non-intersecting and non-overlapping loops on the honeycomb 
lattice \cite{DMNS}. This loop gas representation enables further
mappings on the Kagom\'e 6-vertex model and the Coulomb gas, and has
therefore played a crucial role as a step in the derivation of exact
results for the honeycomb O($n$) model \cite{N,Baxter}.

Just as for the Potts model, tricriticality can be induced by
introducing a sufficient number of vacancies. This was already
known \cite{DS,BBN} for the case $n=0$ which describes the collapse
of a polymer at the so-called theta point, induced by attractive
interactions between the polymer segments, and for the Ising case
$n=1$ \cite{Blume,Capel} where the existing results for the 
tricritical $q=2$ Potts model apply \cite{BN2}.
For the O($n$) loop model, the existence of tricritical points was 
further confirmed by transfer-matrix analyses  for a range of values
of $n$ \cite{GBN,GBL}. Whereas this work yielded reasonably accurate
values for some universal parameters, no exact
formulas were found for these parameters as a function of $n$.

Recently Janke and Schakel \cite{js} reconsidered the numerical
evidence for the tricritical O($n$) model and postulated that the
conformal classification in terms of the Kac formula \cite{FQS,Kac}
of the magnetic exponent, which is known to be $p=q=m/2$ for $n=0$ and 1,
generalizes to other $n$.

In order to find the `missing link', which is the relation between $n$
and the conformal anomaly $c$, a clue is provided by the observation
that a {\em critical} O($n$) model corresponds with a {\em tricritical}
$q=n^2$-state Potts model \cite{N}. It would thus be interesting to
bring the tricritical Potts model into an even higher critical state.
This appears to be possible \cite{NWB,KBN} by the simultaneous 
introduction of vacancies and their dual counterparts, four-spin
couplings, into the Potts model. The equivalent loop model on
the surrounding lattice then appears to have a parameter subspace
where the Yang-Baxter equations are solvable \cite{NWB}. Out of
four branches of solutions parametrized by $q$, one was identified
as a branch of tri-tricritical Potts transitions \cite{NWB,KBN}.
The conformal anomaly and exponents of this model are found as a 
function of $q$ by using an alternative representation as a
Temperley-Lieb model \cite{NWB}.

Since the equivalent loop model has loop weight $\sqrt{q}$, it
seems an appealing possibility that the universal properties of
the $q$-state tri-tricritical Potts model match those of the
tricritical O($n=\sqrt{q}$) loop model. The conformal anomaly derived
in Ref.~\onlinecite{NWB} is, expressed in $n=\sqrt{q}$, determined
by the following equations:
\begin{equation}
c=1-\frac{6}{m(m+1)}, \mbox{\hspace{3mm}} 2\cos\frac{\pi}{m+1}= \Delta,
\mbox{\hspace{3mm}}  \Delta -\frac{1}{ \Delta} =n
\label{ca}
\end{equation}
Furthermore, Ref.~\onlinecite{NWB} yielded scaling dimensions of which
we quote three as
\begin{equation}
X_j=(k_j^2-1)/[2m(m+1)], 
\label{xj}
\end{equation}
where we introduced an index $j=1$, 2 or 3, and $k_j$ is given by
$\Delta_j=2 \cos [k_j\pi/(m+1)]$, with 
$\Delta_1=1/\Delta$, $\Delta_2=-1/\Delta$, and $\Delta_3=-\Delta$.

In order to verify the relation with the tri-tricritical Potts model,
we employ transfer-matrix
calculations for the loop model on the honeycomb lattice with
vacancies. These are introduced as face variables located on the
elementary hexagons. They have two possible states: vacant (weight
$v$) or occupied (weight $1-v$). Furthermore there is an $n$-component
vector spin $\vec{S}_{i}$ on each vertex $i$ that is surrounded by
three occupied hexagons. The one-spin weight distribution is 
isotropic and normalized such that $\int d{\vec{S}}=1$ and
$\int d{\vec{S} \vec{S}\cdot \vec{S}}=n$.
The partition function  given by \cite{N} thus generalizes to
\begin{equation}
Z=\sum_{\cal L}v^{N_{v}}(1-v)^{N-N_{v}}\int \prod_{i| \cal L} d{\vec{S}}_i
\prod_{\langle ij \rangle} (1+w{\vec{S}}_i \cdot {\vec{S}}_j)
\label{Zspin}
\end{equation}
where the sum is on all configuration variables: $\cal L$ is a subset of
the dual lattice and represents the occupied faces of the honeycomb lattice.
The product over $i |\cal L$ includes all vertices except those of the
vacant hexagons, $N_v$ is the number of vacant faces, $N$ is the total
number of faces, $w$ controls the strength of the spin-spin coupling,
and $\langle ij \rangle$ denotes all nearest-neighbor spin pairs.
The mapping on a loop model proceeds along the same lines as in
Ref.~\onlinecite{DMNS} and leads to the following partition sum
\begin{equation}
Z=\sum_{\cal L}\sum_{\cal G|L}v^{N_{v}}(1-v)^{N-N_{v}}w^{N_{w}}
n^{N_{l}}
\label{Zloop}
\end{equation}
where the first sum is on all possible configurations ${\cal L}$ of
occupied faces, and the second one over all configurations ${\cal G}$
of closed loops on the honeycomb lattice that avoid the empty faces;
$N_{w}$ is the number of vertices on ${\cal G}$, and $N_l$ the
number of loops. 

The transfer matrix is constructed for a model wrapped on a cylinder,
whose axis is parallel to one of the lattice edge
directions. The unit of length is defined as the small diameter of an
elementary hexagon. The transfer matrix keeps track of the change of
the number of loops, vacancies, and visited vertices when a new layer
of sites is added to the cylinder. Its largest eigenvalue
determines finite-size data for the free energy density, from which
the conformal anomaly can be estimated \cite{BCN}.
Three more eigenvalues $\lambda_i$ were calculated. These determine
three correlation lengths and allow finite-size estimates $X_i(v,w,L)$
of the corresponding scaling dimensions $X_i$ \cite{Cardy-xi}.
The temperature dimension $X_t$ was estimated from the second
eigenvalue of the transfer matrix, and the magnetic dimension $X_h$
from a modified transfer matrix with a single loop segment running in
the length direction of the cylinder. The `interface' exponent $X_m$
was estimated by inserting a column with bond weights of the opposite
sign. Further details about the transfer-matrix technique are given in
Refs.~\onlinecite{GBN,GBL,BN}. 

We parametrize the vicinity of the tricritical fixed point by two
relevant temperature-like fields $t_1$ and $t_2$, and by an irrelevant
field $u$. The associated exponents are $y_{t_1}$, $y_{t_2}$ and
$y_u$ respectively, with $y_{t_1}>y_{t_2}$.

The tricritical point is estimated by simultaneously solving the unknowns 
$v$ and $w$ in the two equations
\begin{equation}
X_i(v,w,L)=X_i(v,w,L-1)=X_i(v,w,L-2)
\label{2eq}
\end{equation}
where the functions $X_i$ ($i=h,t,m$) are provided by the transfer-matrix
algorithm. Expansion of the finite-size-scaling function in the vicinity
of the tricritical point yields that the solution $v(L)$ of Eq.~(\ref{2eq})
converges to the tricritical value $v^{\rm (tri)}$ of $v$ as
\begin{equation}
v(L)=v^{\rm (tri)}+ a L ^{y_u-y_{t_2}} + \cdots
\end{equation}
where $a$ is an unknown constant; and $w(v,L)$ similarly
converges to the tricritical value $w^{\rm (tri)}$.
The values $X_i(L)$ taken at the solutions of Eq.~(\ref{2eq})
converge to the tricritical scaling dimension $X_i$ as
\begin{equation}
X_i(L)=X_i + b u L^{y_u} +\cdots
\label{xsol}
\end{equation}
where $b$ is another unknown constant.
We applied this procedure both for $X_i=X_h$ and for $X_i=X_m$ to
locate the tricritical points and to estimate 
the tricritical exponents from Eq.~(\ref{xsol}). 
These calculations were performed along the same lines as in
Ref.~\onlinecite{GBL}, but here we use larger finite sizes up to $L=14$,
and moreover we include several values for $-2\leq n<0$. 
The results for the tricritical points are listed in Table I, together
with the estimated error margins.
\begin{table}
\caption{Tricritical points as determined from the transfer-matrix
calculations described in the text. The estimated numerical uncertainty
in the last decimal place is shown in parentheses.}
\label{tab_1}
\begin{center}
\begin{tabular}{||l|l|l||}
\hline
$n$  &    $v$        &   $w$         \\   
\hline
-2.0 & 0.3503    (1) & 0.8156    (1) \\
-1.75& 0.3649    (1) & 0.8330    (1) \\
-1.50& 0.380814  (1) & 0.852082  (1) \\
-1.25& 0.3979352 (1) & 0.8726404 (1) \\
-1.00& 0.4163568 (3) & 0.8949268 (1) \\
-0.75& 0.4360088 (1) & 0.9189617 (2) \\
-0.50& 0.4566834 (2) & 0.9446100 (2) \\
-0.25& 0.4781475 (2) & 0.9717428 (2) \\
0    & 1/2           & 1             \\
0.25 & 0.521805  (1) & 1.028950  (1) \\
0.50 & 0.54313   (1) & 1.05812   (1) \\
0.75 & 0.56361   (2) & 1.08708   (2) \\
1.00 & 0.5830    (1) & 1.1155    (1) \\
1.25 & 0.6010    (1) & 1.1429    (1) \\
1.50 & 0.6175    (1) & 1.1688    (1) \\
1.75 & 0.6321    (1) & 1.1928    (1) \\
2.00 & 0.6452    (1) & 1.2145    (1) \\
\hline
\end{tabular}
\end{center}
\end{table}
The analyses using $X_h$ and $X_m$ generated consistent results and 
allowed us to check the numerical uncertainties.

We also obtained finite-size estimates of $X_t$ at the tricritical 
points thus calculated, and extrapolated these data.
The results for the conformal anomaly and the three exponents are
listed in Table II, together with the estimated error margins.

\begin{table}
\caption{Conformal anomaly and tricritical exponents as determined from
the transfer-matrix calculations described in the text. Estimated error
margins in the last decimal place are given in parentheses. The numerical
results are indicated by `(num)'. For comparison, we include theoretical
values obtained from Eqs.~(\protect\ref{ca}), (\protect\ref{xj}), and 
(\protect\ref{Kac}).
For $n<-3/2$, the temperature exponent $X_t$ becomes complex.
}
\label{tab_2}
\begin{center}
\begin{tabular}{||l|l|l|l|l||}
\hline
 $n$ &$c$ (num)     &$c$ (exact)  &$X_m$ (num)   &$X_m$ (exact) \\
\hline
$-2.0 $&$-0.9914 $ (2)&$-0.9915599$&$-0.202   $ (1)&$-0.2017990$ \\
$-1.75$&$-0.9108 $ (2)&$-0.9109986$&$-0.1765  $ (2)&$-0.1769723$ \\
$-1.50$&$-0.8196 $ (2)&$-0.8197365$&$-0.15166 $ (3)&$-0.1516447$ \\
$-1.25$&$-0.7164 $ (1)&$-0.7164556$&$-0.12596 $ (3)&$-0.1259301$ \\
$-1.00$&$-0.6000 $ (1)&$-6/10     $&$-0.10001 $ (2)&$-1/10     $ \\
$-0.75$&$-0.46962$ (1)&$-0.4696195$&$-0.07410 $ (1)&$-0.0740955$ \\
$-0.50$&$-0.32528$ (1)&$-0.3252829$&$-0.04853 $ (1)&$-0.0485319$ \\
$-0.25$&$-0.16799$ (1)&$-0.1679953$&$-0.023691$ (1)&$-0.0236917$ \\
  0    &  0           &  0         &  0            &  0          \\
  0.25 &  0.17526 (1) &  0.1752630 &  0.022111 (1) &  0.0221110  \\
  0.50 &  0.35348 (1) &  0.3534792 &  0.04224  (1) &  0.0422357  \\
  0.75 &  0.52994 (1) &  0.5299489 &  0.05999  (1) &  0.0600004  \\
  1.00 &  0.70000 (1) &  7/10      &  0.0749   (1) &  3/40       \\
  1.25 &  0.860   (1) &  0.8589769 &  0.0867   (2) &  0.0865052  \\
  1.50 &  1.001   (2) &  1         &  0.094    (5) &  0.0880192  \\
  1.75 &  1.04    (4) &            &  0.098    (5) &             \\
  2.00 &  1.05    (2) &            &  0.10     (1) &             \\
\hline
   $n$ &$X_t$ (num)   &$X_t$ (exact)&$X_h$ (num)   &$X_h$ (exact)\\
\hline
$-2.0 $& ~~~~------   & ~~~~------ &$-0.094   $ (1)& $-0.0951627$\\
$-1.75$& ~~~~------   & ~~~~------ &$-0.087   $ (1)& $-0.0876431$\\
$-1.50$& 0.709    (1) & 0.7097847  &$-0.0792  $ (1)& $-0.0790909$\\
$-1.25$& 0.4817   (2) & 0.4814739  &$-0.0694  $ (1)& $-0.0693653$\\
$-1.00$& 0.4000   (2) & 2/5        &$-0.0584  $ (1)& $-7/120    $\\
$-0.75$& 0.3445   (2) & 0.3446681  &$-0.04593 $ (3)& $-0.0458895$\\
$-0.50$& 0.30390  (2) & 0.3039309  &$-0.03199 $ (1)& $-0.0319828$\\
$-0.25$& 0.273220 (1) & 0.2732199  &$-0.016645$ (1)& $-0.0166435$\\
  0.00 & 1/4          & 1/4        & 0             &  0          \\
  0.25 & 0.232500 (1) & 0.2324957  & 0.017731 (1)  &  0.01772952 \\
  0.50 & 0.2193   (1) & 0.2192386  & 0.03628  (1)  &  0.03627658 \\
  0.75 & 0.2090   (2) & 0.2088741  & 0.05539  (1)  &  0.05539746 \\
  1.00 & 0.2000   (1) & 1/5        & 0.07500  (2)  &  3/40       \\
  1.25 & 0.193    (1) & 0.1906800  & 0.0950   (2)  &  0.09549714 \\
  1.50 & 0.180    (5) & 0.1684499  & 0.12     (1)  &  1/8        \\
  1.75 & 0.183   (10) &            & 0.13     (1)  &             \\
  2.00 & 0.184   (10) &            & 0.15     (2)  &             \\
\hline
\end{tabular}
\end{center}
\end{table}

\begin{figure}
\includegraphics{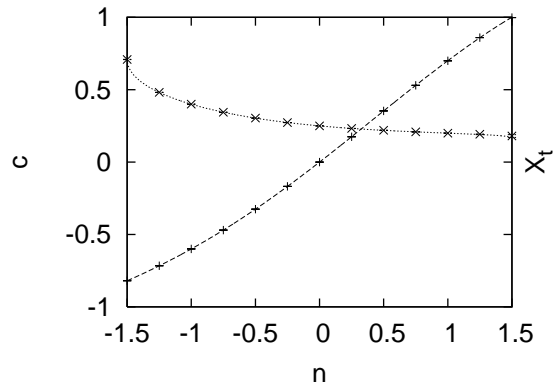}
\caption{Conformal anomaly ($+$) and temperature exponent
($\times$) of the tricritical O($n$) model
vs. $n$. The data points show the numerical data, the curves the
theoretical predictions.}
\label{ccn}
\end{figure}

\begin{figure}
\includegraphics{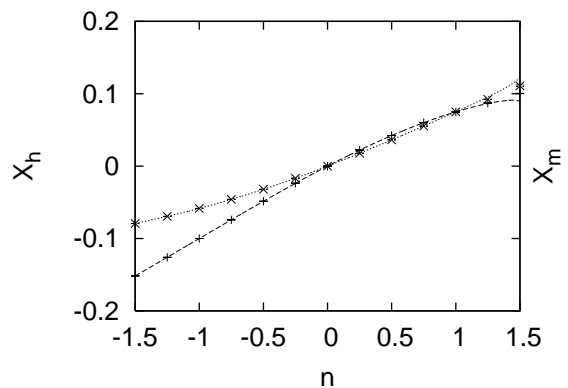}
\caption{Scaling dimensions $X_h$ ($\times$) and $X_m$ ($+$) of the
tricritical O($n$) model vs. $n$. The data points show the numerical
data, the curves the theoretical predictions. }
\label{expn}
\end{figure}

A comparison of the numerical results for the conformal anomaly with
Eq.~(\ref{ca}), as given in Table \ref{tab_2}, strongly supports the
classification of the tricritical O($n$) model as proposed for
$n\leq 3/2$. Our numerical results for $X_t$ agree with $X_2$ in
Eq.~(\ref{xj}), those for $X_m$ agree with $X_1$. 
Using the value of the conformal anomaly and $m$ as a function $n$,
we confirm that the numerical results for the magnetic scaling
dimension agree with the entry ($p=m/2, q=m/2$) in the Kac formula
\begin{equation}
X_{p,q}=\frac{[p(m+1)-qm]^2-1}{2m(m+1)}, 
\label{Kac}
\end{equation}
The $n>0$ results correspond with branch 1 as defined in
Ref.~\onlinecite{NWB}, and those for $n<0$  with branch 2.
The numerical results and theoretical values of the conformal anomaly 
and  the dimensions are shown as function of $n$ in Figs.~(\ref{ccn})
and (\ref{expn}).

The expressions for $X_1$ and $X_2$ in Eq.~(\ref{xj}) are not reproduced
by entries in the Kac table, at least not with index pairs that are
linear in $m$. This made it difficult to conjecture the exact values
of $X_m$ and $X_t$ from numerical data alone, even if supported by
data for $c$. This problem did not apply to $X_h$ which 
appears in the Kac table.  

Remarkably, the Potts tri-tricritical branch ends at $q=9/4$. For
$q>9/4$ the model is no longer critical and the transition
probably turns first-order \cite{NWB,KBN}. The equivalence $q=n^2$
thus yields the result that the tricritical O($n$) branch ends at
$n=3/2$, possibly with a discontinuous transition for $n>3/2$.
At first sight, the numerical results for $3/2<n \leq 2$ may not
seem suggestive of a discontinuous transition, and allow only a
very weak discontinuity. But it is clear from Tables I and II
that the estimated errors tend to increase with $n$ for
$n\gae 3/2$, as a result of deteriorating finite-size convergence.
This is consistent with the possibility that an operator becomes
marginal at $n=3/2$, in line with $c=1$ (see Table \ref{tab_2}). 
Therefore, one may expect similar phenomena
as for the $q > 4$ Potts model, where the marginal
operator leads to misleadingly slow finite-size convergence which
obscures the weak first-order character in a range of $q$ near 4.

The results presented above apply to the non-intersecting loop model.
Loop intersections are irrelevant in the critical O($n$) model,
but they are relevant in the low-temperature phase \cite{jrs}.
While the possible relevance of such intersections could modify the
universal behavior, this appears not to be the case
for the $n=1$ tricritical O($n$) loop model,
since its exponents agree with those of the corresponding spin model.

The scenario sketched above indicates that the critical and tricritical
O($n$) branches are {\em not connected}, and does not provide a relation   
between the tricritical O($n$) model and the critical Potts model,
such as was recently suggested \cite{js}.

This research is supported by the National Science Foundation of
China  under Grant \#10105001, by a grant from Beijing Normal
University, and, in part, by the FOM (`Fundamenteel Onderzoek der
Materie') Foundation of the Netherlands.

\end{document}